\documentclass[12pt,preprint]{aastex}

\shorttitle{optical and high energy emission from AXPs}

\begin{document}

\title{On the Infrared, Optical and High Energy Emission 
\\from the  
Anomalous X-Ray Pulsar 4U 0142+615}


\author{\"{U}. Ertan\altaffilmark{1} and 
K. S. Cheng\altaffilmark{2}}

\affil{Department of Physics, The University of
Hong Kong, Hong Kong, China}

\altaffiltext{1}{Also, Sabanc{\i} University,
Orhanl{\i}$-$Tuzla 34956, {\.I}stanbul/TURKEY,  
unal@sabanciuniv.edu}
\altaffiltext{2}{hrspksc@hkucc.hku.hk} 

\begin{abstract}
We show that the observed {\it pulsed} optical emission of the 
anomalous X-ray pulsar 4U 0142+61 can be accounted for 
by both the magnetar outer gap models and the disk-star 
dynamo gap models, therefore is not an evidence in favor 
of one of these models as its responsible mechanism. 
Nevertheless, the estimated high energy gamma-ray spectra 
from these models have different power-low indices, and can be tested 
by future observations of the Gamma-ray Large-Area 
Space Telescope (GLAST).    
Furthermore, we show by analytical estimations that   
the expectations of a standard  disk model is in agreement 
with the observed {\it unpulsed} optical and 
infrared  luminosities of the AXP 4U 0142+61.    
 
\end{abstract}

\def\la{\raise.5ex\hbox{$<$}\kern-.8em\lower 1mm\hbox{$\sim$}}
\def\ga{\raise.5ex\hbox{$>$}\kern-.8em\lower 1mm\hbox{$\sim$}}
\def\be{\begin{equation}}
\def\ee{\end{equation}}  
\def\ba{\begin{eqnarray}}
\def\ea{\end{eqnarray}}  
\def\be{\begin{equation}}
\def\ee{\end{equation}}  
\def\ba{\begin{eqnarray}}
\def\ea{\end{eqnarray}}  
\def\m{\mbox}
\def\O{\Omega}
\def\d{\partial}
\def\a{\alpha}
\def\Mdot*{\dot{M}_*}
\def\Mdotin{\dot{M}_{\mbox{in}}}
\def\Lin{L_{\mbox{in}}}
\def\Lx{L_{\mbox{\scriptsize x}}}
\def\LR{L_{\mbox{\scriptsize R}}}
\def\AR{A_{\mbox{\scriptsize R}}}
\def\AK{A_{\mbox{\scriptsize K}}}
\def\Ks{K_{\mbox{\scriptsize s}}}
\def\LIR{L_{\mbox{\scriptsize IR}}}
\def\Teff{T_{\mbox{\scriptsize eff}}}
\def\LK{L_{\mbox{\scriptsize K}}}
\def\Rin{R_{\mbox{\scriptsize in}}}
\def\Rout{R_{\mbox{\scriptsize out}}}
\def\Ldisk{L_{\mbox{\scriptsize disk}}}
\def\dEb{\delta E_{\mbox{burst}}}
\def\dEx{\delta E_{\mbox{x}}}
\def\Bb{\beta_{\mbox{b}}}
\def\Be{\beta_{\mbox{e}}}
\def\dMin{\delta M_{\mbox{in}}}
\def\dM*{\delta M_*}
\def\r_c{\r_{\mbox{co}}}
\def\r_0{\r_{\circ}}
\def\rl{\r_{\m{\scriptsize l}}}
\def\p{\propto}
\def\ss{\scriptsize}
\def\s{\small}
\def\O{\Omega}




\section{Introduction}

Anomalous X-ray pulsars (AXPs) are neutron stars (NSs) which are characterized 
by their persistent X-ray luminosities 
($\Lx \sim 10^{34} - 10^{36}$ ergs s$^{-1}$) which are 
well above their rotation powers. Observations show no evidence 
for the existence of massive companions. Their periods are 
clustered in a remarkably 
narrow range $P\sim 6 - 12$ s.       
Some of the AXPs belong to SNRs which show that they are young objects 
(See Mereghetti et al. 2002, for a review of AXPs). The X-ray
luminosities, 
periods, and black body temperatures of AXPs are similar to those of 
soft gamma-ray repeaters (SGRs) (see Hurley 2000 for a review). 
Furthermore, recently reported SGR-like bursts from two of the 
AXPs (Kaspi $\&$ Gavriil 2002, Gavriil et al. 2002) 
imply that these sources probably have similar origins. 

Magnetar models (Dunacan $\&$ Thompson 1992, Thompson $\&$ Duncan 1995) 
are successful in explaining the fluences and the luminosities of the 
normal ($L~\la~10^{42}$ erg s$^{-1}$) and the giant 
($L~\ga~10^{44}$ erg s$^{-1}$) super-Eddington bursts of SGRs. In
the 
magnetar models, surface magnetic fields are $\ga~10^{14}$ G, and   
huge magnetic energy release through instabilities from inside the NS 
is the source of these bursts.  The persistent thermal X-ray emission 
was suggested to be due to the surface heating by magnetic field decay, 
while the accelerated particles in the magnetosphere by the 
Alfv$\acute{\m{e}}$n waves  produced by the small scale fractures on the
NS 
surface could be the source of the non-thermal emission 
(Thompson $\&$ Duncan 1996).  Spin-down torque is provided by the 
magnetic {\it dipole} radiation. Despite the success of the 
magnetar model in explaining  the burst energetics, it can account 
for the period clustering of SGRs and AXPs only in one set of decay models 
with special conditions  (Colpi, Geppert $\&$ Page 2001). 

On the other hand, fall-back disk models, which have difficulties in 
explaining the short time scales of the super-Eddington bursts, 
can account for the period clustering and the persistent 
X-ray emission  of these sources (Chatterjee, Hernquist $\&$ 
Narayan 2000; Alpar 2001; Marsden, Lingenfelter $\&$ Rothschild 2001). 
The enhancement phase of the persistent X-ray emission 
following the giant flare 
of the SGR 1900+14 can be explained by the post-burst release 
of the inner disk which has been pushed back by the burst energy 
(Ertan $\&$ Alpar 2003)
(see also Lyubarsky, Eichler $\&$ Thompson (2002) for an explanation 
based on crust cooling after being heated by the burst energy). 
Moreover, it was shown that the young
NSs with fall-back
disks can produce AXPs within the time scales of the
ages of their associated supernova remnants with reasonable
parameters (Chatterjee $\&$ Hernquist 2000; Ek\c{s}i $\&$ Alpar 2003a,b). 
   
Magnetar models have no detailed predictions for the IR and optical 
emissions seen from the AXPs which, in this work, is one of our main
concerns based on the disk models.
Out of five observed AXPs (and two candidates), four sources 
were reported to have IR counterparts 
(Kaspi et al. 2002, Israel et al. 2002, Israel et al. 2003).  
 One of them 
(4U 0142+61) was also detected in the R-band 
(Hulleman et al. 2000).
The optical/IR luminosities of these AXPs lie 
above the extrapolation of the black body components 
of their X-ray spectra 
(Hulleman et al. 2001, Wang $\&$ Chakrabarty 2002, Israel et al. 2003)
indicating different origins for the optical/IR emission, and the thermal 
X-rays likely to be emitted near the NS surface. 
The observed ratio of the total R-band luminosity       
to the X-ray luminosity is $\sim 1.4\times 10^{-4}$  for  AXP 4U 0142+61  
(Hulleman et al. 2000, Hulleman 2001), and was claimed 
to be too low for the origin of the optical light to be 
an accretion disk.
Kern and Martin (2002) reported that the optical (in the R-band) 
emission from the AXP 4U 0142+61 has a pulsed component with 
a pulsed fraction $h_{\m{\ss R}} \simeq 0.27$ which is much 
higher than the  X-ray pulsed fraction $\sim 5 \% ~(1-2$ keV), 
which strongly suggests  different emission sites for the pulsed 
X-ray and optical emissions. It was argued that the pulsed optical 
radiation is an indication that AXPs are magnetars.

In present work, we show that the pulsed optical emission 
can be accounted for by both 
the disk-star dynamo gap model (Cheng $\&$ Ruderman 1991) and 
the magnetar outer gap model
(Cheng $\&$ Zhang 2001),  and does not exclude one or the 
other (Sections 2 $\&$ 3). In Sec. 4, we show that the 
two models can be observationally tested through their expected different 
high energy gamma-ray spectral characteristics. 
Furthermore, in Sec. 5,  we show by analytical estimations that the
R and K-band luminosities  
expected from a standard disk  model 
is in agreement with the observations of the AXP 4U 0142+61.   
 
\section{Disk-Star Dynamo Model}

For simplicity, we consider a NS dipole magnetic field 
aligned with the spin axis.   
When the inner rim of a Keplerian disk rotates faster than 
the neutron star,  
a gap (accelerator) is formed around the null surface ({$\bf \O.B$}=0)
separating the charges in the magnetosphere.
The electromotive force of the disk-star dynamo 
(see Cheng and Ruderman (1991) for details, CR thereafter) 
is given by
\be
{\cal E}(r)=\int^{r}_{r_0} ~(\O_{\m{\ss K}}(r) - \O_*) ~B_z(r) 
\frac{r dr}{c}   
\label{1}
\ee
where $B_z(r)$ is the vertical component of the magnetic field 
through the disk, $\O_{\m{\ss K}}(r) $ 
the local Keplerian angular velocity 
of the disk, $\O_*$ the angular frequency of the NS, 
and ${r_0}$ the inner radius of the disk.  
In a stationary gap, the electromotive force is balanced by the 
potential drop $V_{\m{\ss g}}$ across the gap . Substituting 
$\O_{\m{\ss K}}(r) = (GM/r^3)^{1/2}$ in Eq.1, 
and neglecting the effects of inflowing disk matter,  we obtain 
\be
{\cal E}(r) = R_*^3 B_* \left[\frac{2}{5}
\left(\frac{GM_*}{c^2}\right)^{1/2} 
\left(\frac{1}{r_0^{5/2}} - \frac{1}{r^{5/2}} \right) -
\frac {\O_*}{c} 
\left(\frac{1}{r_0} - \frac{1}{r} \right)
\right] \label{2}
\ee
where  
$R_*$, $B_*$, and  $M_*$ are the radius, surface magnetic field and 
the mass of the NS respectively. 
Whether the gap is closed or open depends on 
how close the co-rotation radius 
$r_c = (GM/\O_*^2)^{1/3}$ is to the inner disk radius $r_0$. 
If $r_0$ is less than $1.8~ r_c$ the gap cannot close 
(the geometry of the open and closed gaps
are given in Figs. 2 $\&$ 3 of CR).   
In the limit $\O_* \ll \O_{\m{\ss K}}(r_0)$ 
the potential drop in Eq.(\ref{2}) 
can be as high as 
\be
{\cal E}(r) \sim 4\times 10^{14}~ r_{0,8}^{5/2}~ B_{*,12}~ 
\left(\frac{M}{M_{\odot}}\right)^{1/2}~ R_{*,6}^3~ 
\left[1-\left(\frac{r_0}{r}\right)^{5/2}\right] 
~  {\m{~V}}     
\label{3}
\ee
where we use the notation $A= A_{,i}\times 10^i$. 
This potential drop can only be reached by a static gap 
in which the pair production is negligible. 
In a more realistic stationary gap, the potential drop 
$V_{\m{\ss g}}$ is limited by the flow of electrons and positrons 
through the gap, especially when there is copious amount 
of X-ray photons traversing the gap. The X-rays are up-scattered
by the inverse Compton collisions with the gap accelerated 
electrons and positrons. The gamma-rays produced by these 
inverse Compton scatterings create secondary $e^\pm$ pairs 
through the collisions with the X-rays. These secondary pairs 
emit synchrotron emission out of the gap, which 
we test whether to be responsible for the observed 
pulsed optical emission.  
For such a pair production limited gap, the maximum potential 
drop can be estimated as
\be
V_{\m{\s g}} = 6 \times 10^{10}~ L_{{\m{x}},36}~ 
\left(\frac{\langle E_{\m{x}} \rangle}{1 {\m{keV}}}\right)^{-1}~ 
B_{*,12}^{-1/2}~ R_{*,6}^{-1}~ {\m{V}}
\label{4}
\ee  
(CR). The energies of the primary pairs accelerated in this 
potential drop $V_{\m{\ss g}}$ are limited by their collisions 
with the X-ray photons.   
On the other hand, the gap power is limited by the power 
of the inflowing disk matter from $r_c$ 
down to the inner disk radius $r_0$,  
\be
P\sim G M \dot{M}~ \left(\frac{1}{r_0} - \frac{1}{r_c}
\right).  \label{5}  
\ee
The inner disk radius can be written as 
\be
r_0 \simeq \beta \frac{r_{\m {\ss A}}}{2} 
=2.9 \times 10^8 ~\beta ~L_{{\m{\s x}},36}^{-2/7}~ B_{*,12}^{4/7}~ 
\left(\frac{M_*}{M_{\odot}}\right)^{1/7}~ R_{*,6}^{10/7} ~{\m{cm}}
\label{6}
\ee
(Ghosh $\&$ Lamb 1979) where $r_{\m {\ss A}}$ is the 
Alfv$\acute{\m{e}}$n radius and $\beta\sim 1$. 
For the AXP 4U 0142+61,   
$r_c \sim 7\times 10^8$ cm, and 
the X-ray luminosity is 
$L_{\m{\ss x}}=6.5\times 10^{35}$~ ergs s$^{-1}$ for a
source distance $d= 3$ kpc. 
$d$ was reported to be greater than 2.7 kpc
(Hulleman et al. 2000). The accretion rate is related to the X-ray luminosity 
through $\dot{M} = L_{\m{\ss x}} R_*/GM\sim 3.4\times 10^{15}$ g s$^{-1}$. 
Then, Eqs. (5) $\&$ (6) give a gap power
$P_{\m{\ss g}}\sim 1\times 10^{33}$ ergs s$^{-1}$.  
The fraction of the
gap power that is taken out of the gap by the secondary pairs 
depends on the pair production optical depth of the X-rays to 
primary gamma-rays. The density of the thermal X-ray photons  
at $r\sim r_0 $ can be estimated as
\be
\langle n\rangle~ \sim \frac{L_x}{4 \pi 
(r_0)^2~ c 
\langle E_{\m{\ss x}} \rangle} \sim  10^{16}~{\m{cm}}^{-3}    
\label{7}
\ee
where $\langle E_{\m{\ss x}} \rangle \approx 3 kT$ is the typical  
energies of the X-rays. For 4U 0142+61, $3 kT\approx 1.2$~keV 
(Israel et al. 1999, Paul et al. 2000,
Mereghetti et al 2002, Juett et al. 2002). 
The optical depth for the pair production of 
the primary gamma-rays with the soft X-rays becomes  
\be  
\tau \sim~ \langle n \rangle~ \sigma_{\m{\ss p}}~ l~  
\sim(10^{16} {\m{cm}}^{-3})~(10^{-25}~{\m{cm}}^2) 
~(5\times10^8 {\m{cm}})
\sim 0.5
\label{8}  
\ee
where $\sigma_{\m{\ss p}} \approx 10^{25}$ cm$^2$ is the 
cross section for the pair production, and $l \sim r$ is the length of the 
gap along the field lines ($l$ can exceed $r$ considerably 
due to the azimuthal winding of the magnetic field lines 
across the gap). 
Eq.(\ref{8}) implies that   
about half of the gap power is transferred to the 
secondary $e^\pm$ pairs. We investigate whether the synchrotron emission 
from these secondary pairs just out of the gap can   
account for the observed optical pulsed emission from 
AXP 4U 0142+61. 

Energy distribution  of the secondary $e^\pm$ pairs can 
be approximated as
\be
N_e \sim \gamma^{-2}~ \ln\left(\frac{\gamma_{\m{\ss max}}}
{\gamma}\right)
\label{9}
\ee
(Cheng, Ho $\&$ Ruderman 1986) where 
$\gamma$ is the Lorentz factor of the 
secondary electrons and positrons such that 
$\gamma_{\m{\ss min}}< \gamma< \gamma_{\m{\ss max}}$. 
The synchrotron energy spectrum 
from these secondary  pairs is
given by 
\be
{\it f(E_{\gamma})} \p \int^{\gamma_{max}}_{\gamma_{min}} 
d\gamma~ N_e(\gamma) F(x) \label{10}
\ee
where $F(x)= x \int^{\infty}_{x} K_{5/3}(y) dy$, 
$K_{5/3}$ is the modified Bessel function, 
$x= E_{\gamma} / E_{\m{\ss syn}}$, and 
$E_{\m{\ss syn}}\simeq (3/2)~ \gamma^2  \hbar (e B/ m_e c)$. 
The observed energy spectrum becomes 
\be
{\cal F}(E_{\gamma}) \approx \frac{1}{\Delta\Omega~ d^2} 
~{\it f}(E_{\gamma})
\label{11}
\ee
where $\Delta\Omega~$ is the beaming solid angle of 
the synchrotron emission, and $d$ is the distance to the 
source. The total synchrotron luminosity 
$L_{\m{\ss syn}}$ is found 
by integrating {$\it f(E_{\gamma})$} overall frequencies. 
The lower cut-off of the spectrum is determined by the 
minimum Lorentz factor $\gamma_{\m{\ss min}}$ of the secondary pairs 
for which we take $\gamma_{\m{\ss min}}= 1$. 
The maximum Lorentz factor of the secondary pairs 
is $\gamma_{\m{\ss max}}\sim$ MeV$/\langle E_{\m{\ss x}}\rangle$ 
$\sim$ few $\times 10^3$
(Cheng, Ho, $\&$ Ruderman 1986; Cheng $\&$ Ding 1994). The exact position
of 
$\gamma_{\m{\ss max}}$ does not affect the results significantly, 
we take $\gamma_{\m{\ss max}}= 3\times 10^3$. 
The synchrotron radiation is emitted at a radial distance 
$r~\ga ~r_0$ where the magnetic field strength reduces to about 
$\sim 3\times 10^4$ G for a dipole field with $B_* \sim 10^{12}$ G 
on the surface of the NS. The resultant photon spectrum of 
the synchrotron emission is given in Fig. 1.   

Our aim is to compare the model pulsed optical luminosity in the  
R-band with that of the observations.
The ratio of the observed  optical 
luminosity in the R-band ($\sim 1.7 - 2.1$ eV) to the 
X-ray luminosity was reported to be 
~$\approx 1.4\times 10^{-4}$ from the unabsorbed 
fluxes (Hulleman et al. 2000) with a pulsed fraction of the 
optical light $h_{\m{\ss R}} \sim 0.3$ (Kern $\&$ Martin 2002). Then,  
the observed ~{\it pulsed}~ R-band luminosity 
is $L_{\m{\ss R}}^{\m{\ss obs}} \sim 2.5\times 10^{31}~(\Delta\O/4\pi) $ 
ergs s$^{-1}$~ for 
$L_{\m{\ss x}}\simeq 6\times 10^{35}$ ergs s$^{-1}$ where 
$\Delta\O$ is the beaming angle of the synchrotron emission.   
To compare our results with the 
observations, we calculated the ratio of the 
synchrotron R-band luminosity  
to the total synchrotron luminosity and find that 
$L_{\m{\ss R}}/L_{\m{\ss syn}}\approx 1\times 10^{-2}$.
Since the pair production optical depth provided by the 
soft X-rays to the primary gamma-rays is around one half, 
$L_{\m{\ss syn}} \sim P_{\m{\ss gap}}/2$ 
$\sim 5\times 10^{32}~$ ergs~ s$^{-1}$
for the  AXP 4U 0142+61,  
 which  means  
$L_{\m{\ss R}}\approx 5\times 10^{30}$ ergs~ s$^{-1}$. 
For a beaming angle $\Delta\O \sim 2.5 $, the optical 
pulsed luminosity  
estimated  by the disk-star dynamo model is in agreement  
with the observations.

\section{Magnetar Model}

The possible gamma-ray emission properties from the outer 
gaps for both AXPs (with soft 
thermal X-ray input) and SGRs (including the effect of the 
hard X-ray component) has been worked in detail by assuming         
dipole magnetar fields with $B_*~ \ga~ 10^{14}$ G   
(Cheng $\&$ Zhang 2001,  Zhang $\&$ Cheng 2002). 
In the outer gap models of AXPs, the primary gamma-rays 
are produced by the curvature emission of the accelerated primary
electrons and positrons inside the gap (Cheng $\&$ Zhang 2001), 
which is different from the gamma-ray production mechanism of 
the disk-star dynamo model. 
The primary gamma-ray luminosity, that is, the power provided by 
the magnetar outer gap can be written as
\be
L_{\gamma} \approx 4.0 \times 10^{32}~ 
\left(\frac{f}{0.5}\right)^3~ 
\left(\frac{B_*}{10^{14} {\m{G}}}\right)^2~ 
\left(\frac{P}{6 {\m{s}}}\right)^{-4}~ 
\left(\frac{R_*}{15 {\m{km}}}\right)^6~ {\m{ergs s}}^{-1}
\label{12}     
\ee
(Zhang $\&$ Cheng 1997) where $f$ is the fraction of the 
volume of the outer magnetosphere occupied by the 
outer gap which is given by  
\be
f\approx 0.68~ \left(\frac{p}{6 s}\right)^{7/6}~ 
\left(\frac{B}{10^{14} {\m{G}}}\right)^{-1/2}~ 
\left(\frac{T_s(\theta)}{5\times10^6 
{\m{K}}}\right)^{-2/3}~ 
\left(\frac{R_*}{15 {\m{km}}}\right)^{-3/2}
\label{13}
\ee
where $T_{\m{\ss s}}(\theta)$ is the surface temperature of the magnetar. 
Its dependence on the polar angle $\theta$ with respect to 
the magnetic axis is weak and may be ignored. It should be noted 
that if $f> 1$ an outer gap cannot be formed (Zhang $\&$ Cheng 1997).    
For the AXP 4U 0142+61, the reported period is $p=8.69$~s 
and the surface magnetic field estimated from 
$B=3.3\times 10^{19} (P \dot{P})^{1/2}$~ is around 
~$\sim 1.4 \times 10^{14}$ G 
(Mereghetti et al. 2002), and the 
corresponding fractional size of the outer gap is
$f\approx 0.89$. Then the outer gap power of 4U 0142+61 becomes 
$P_{\m{\ss g}}\sim 1\times 10^{33}$ ~erg s$^{-1}$, which is roughly the 
same as the gap power of the disk-star dynamo model. 
Similar to the disk-star 
dynamo model gap, secondary pairs, produced by the collisions between 
the primary gamma-rays and X-ray photons originating 
from the NS, emit synchrotron radiation out of the gap. 
Unlike the disk-star dynamo gap, the outer 
gap of an isolated NS extends to near the light cylinder radius 
$r_{l}= c/\O_*$ which is $\sim 4\times 10^{10}$
cm for 4U 0142+61.  
Replacing $r_0$ by $\sim r_{\ss l}/3$, where the magnetic field decreases 
to about $10^{2}$~ G, 
and setting $l=r_{\ss l}$ 
in Eqs.(\ref{7} $\&$ \ref{8}), we obtain the pair 
production optical depth $\tau_{\m{\ss p}}\sim
0.03~L_{\m{\ss x,36}}$.    
For the maximum primary gamma-ray energies $\sim$ few GeV 
~(Cheng $\&$ Zhang 2001), 
the maximum Lorentz factor of the secondary pairs are 
$\sim$ few $\times 10^3$. In the calculations, we set 
$\gamma_{\m{\ss min}}=1$ and  
$\gamma_{\m{\ss max}}=3\times 10^3$.  
Following the same numerical calculations as that of 
the disk-star dynamo model, 
we found from the resultant spectrum (Eq.{\ref{10}) that the 
ratio $L_{\m{\ss R}}/L_{\m{\ss syn}}$ is about $3\times 10^{-2}$. 
The total synchrotron luminosity of the secondary pairs 
$L_{\m{\ss syn}} \sim \tau_{\m{\ss p}}$ 
$P_{\m{\ss g}}\simeq 3\times 10^{31}$ 
ergs s$^{-1}$, and the corresponding pulsed R-band luminosity 
is $L_{\m{\ss R}}\sim 3\times 10^{-2}~L_{\m{\ss syn}}\sim 9\times 10^{29}$ 
ergs s$^{-1}$. With a solid angle $\Delta\O \sim 0.5$
of the beamed synchrotron emission , 
magnetar model can also account for the observed pulsed optical 
luminosity of the AXP 4U 0142+61 in the R-band. 
The expected synchrotron photon spectrum is presented in Fig. 1. 
The weakness of the magnetar model for 
AXP 4U 0142+61 is that if the surface magnetic field of this 
source is $\sim 1.4\times 10^{14}$ G (Mereghetti et al. 2002),  
then the fraction $f$ of the outer magnetosphere 
occupied by the outer gap (Eq. 13) exceeds unity 
for $R_*$ less than about  $13.9$ km. 
This would mean that there is no outer gap of 4U 0142+41 at all
(Zhang $\&$ Cheng 1997). 
On the other hand, if the outer gap indeed exists in 
4U 0142+61 this implies that the equation of state (EOS) 
of this source cannot be soft EOS 
(cf. Shapiro $\&$ Teukolsky 1983) with magnetar dipole fields.

The spectral cut-off of the disk-star-dynamo model lies at the 
X-rays, whereas it is at about 10 eV for the magnetar model. 
However, the expected X-ray luminosity from the disk-star dynamo 
model   is much less than the observed pulsed X-ray emission which 
is very probably  emitted near the NS, and is not likely to be 
tested by the observations.  In the next Section, we show that the 
two models could be distinguished through their high energy gamma-ray 
spectral properties.  
  
\section{High Energy Emission from the Magnetar and the 
Disk-Star Dynamo Models}

The estimated high energy (keV - $\sim$ 10 GeV) 
emission spectra of the magnetar and
the disk-star dynamo gap models are different, although their total 
high energy luminosities are similar. The outer gaps of magnetar 
models, which lie close to the light cylinder, are optically thin to the 
inverse Compton scattering of the X-rays by the gap accelerated pairs. 
These pairs emit beamed curvature radiation along the outer gap. The
resultant spectra have been worked for both AXPs and SGRs 
(Cheng $\&$ Zhang 2001, Zhang $\&$ Cheng 2002). The energy distribution of
the primary $e^{\pm}$ pairs in the gap is given by 
\be
\frac{dN}{dE_{\m{\ss e}}} \propto \left(
\frac{\gamma}{\gamma_{\circ}}\right)^{-16/3} 
\label{18}        
\ee
(Zhang $\&$ Cheng 1997) where 
$\gamma_{\circ}\approx 2.8\times10^7 f^{1/2} B_{12}^{1/4}$
$P^{-1/4} (R_*/10$ km)$^{3/4}$.     
The Lorentz factor $\gamma$ is given by 
$\gamma(x)\approx \gamma_{\circ}~x^{-3/4}$ for 
$x_{\m{\ss min}} \le x \le x_{\m{\ss max}}$ where 
$x=s/r_{\ss l}$ and $s$ is the local radius of the 
curvature. For the AXP 4U 0142+61, $\gamma_{\circ}
\sim 5\times 10^7$. The value of $x_{\m{\ss min}}$ depends on the magnetic
inclination angle $\alpha$, and it is   
$\sim 2/3$ for 
$\alpha=45^{\circ}$ 
and $\sim 2/5$ for $\alpha=60^{\circ}$.
$x_{\m{\ss max}}$ could be taken $\sim 2$ 
(Cheng $\&$ Zhang 2001). The primary pairs with
Lorentz factors between $\gamma{\m{\ss min}}$ and 
$\gamma{\m{\ss max}}$ emit curvature radiation with typical energies 
$E_{\gamma}\approx (3/2) \hbar \gamma^3 (c/s)$ ~\la~ GeV. 
Above the characteristic energies spectrum has an exponential cut-off, 
while $I(\nu) \propto \nu^{1/3}$ for 
$E < E_{\gamma_{\m{\ss min}}}$.  
The detailed model spectra of AXPs for different period, magnetic field
and inclination angles are given in Cheng $\&$ Zhang (2001). 

On the other hand, for the disk-star dynamo model, the energies of the 
gap accelerated $e^{\pm}$ pairs are limited by the inverse 
Compton collisions by the X-rays. Thereby produced gamma-ray photons 
create $e^{\pm}$ pairs via collisions with the X-rays. The energy
distribution of the pairs in the gap can be approximated by 
Eq. (\ref{9}). For a power-law energy distribution 
$N_{\m{\ss e}}(\gamma)\propto \gamma^{-m}$ 
of the pairs, the resultant intensity spectrum of the up-scattered 
X-rays can be written as 
$I_{\m{\ss c}}(\nu) \propto \nu^{-p} $ with ~$p=(m-1)/2$. For the pair
distribution
given by Eq. (\ref{9}), we have 
$m~=~2~+~[~\ln (\gamma{\m{\ss max}}/\gamma)~]^{-1}$. 
The variation of the term in square brackets is small along the 
large range of the Lorentz factors. For a variation of  
$\gamma{\m{\ss max}}/\gamma$ from $\sim 5$ to $\sim 10^3$, 
$p$ changes from 0.6 to 0.8. 

We see that the expected high energy gamma-ray spectra of the magnetar 
and the disk-star dynamo models have very different 
spectral power-law indices. The model luminosities 
are high enough such that they 
can be tested by the future gamma-ray mission GLAST observations 
(Cheng $\&$ Zhang 2001).

\section{Unpulsed  Optical/IR Emission} 

Since the unpulsed component of the optical/IR emission is expected
to be isotropic, there is no beaming correction, and therefore 
$(\LR/\Lx)_{\m{\ss observed}} \sim 10 ^{-4}$ for the 
unpulsed emission of the AXP 4U 0142+61. 
The reported IR and optical magnitudes are 
$K \approx 19.4$~ mag  and $R\approx 25$~mag 
(Hulleman et al. 2000, Hulleman 2001). 
The  estimated interstellar extinction in the R-band is 
$\AR=4.4$ mag, then it is expected that $\AK \simeq 0.5$ in the IR band. 
We estimate that the  ratio of the observed K-band flux to that of R-band 
$(\LK/\LR)_{\m{\ss observed}}$ 
is  $\sim 10^{-1}$. Now, we will  compare these observations with the 
expectations of a standard accretion disk model. 

We take $\dot{M}_*= \eta \dot{M}_{\m{\ss in}}$ where $\eta$ is the 
fraction of the disk mass flow rate which is  accreted on to the NS, 
and assume that the mass flow through the inner disk regions  emitting  
in the K and R bands can be represented by a steady state mass flow. 
The local dissipation rate $D(r)$ of the disk at the radial distance
$r$ in steady state is given by
\be
D(r)\simeq \frac{3}{8 \pi} \dot{M}~ \frac{G M}{r^3}.  \label{14} 
\ee
For $\Lx \sim 10^{36}$ ergs s$^{-1}$, the accretion rate onto the 
NS becomes $\dot{M}_* \sim 5\times 10^{15}$ g s$^{-1}$ 
$\sim \eta \dot{M}_{\m{\ss in}}$. We obtain
\be
D(r)\approx 1.2\times 10^{14}~ r_{9}^{-3}~ \eta^{-1} {\m{ergs s}}^{-1} 
{\m{cm}}^{-2}
\label{15}
\ee
where $r_9$ is the radial distance in units of $10^9$ cm. 
The corresponding local effective temperature of the disk becomes
\be
\Teff(r)= \left(\frac{D(r)}{\sigma}\right)^{1/4}
\simeq 3.8\times10^4 ~r_{9}^{-3/4} \eta^{-1/4} 
\label{16}
\ee
where $\sigma$ is the Stefan-Boltzmann constant. The central 
wavelength of the 
R-band is 
$\lambda_{\m{\ss R}}\simeq 138$ nm. The effective temperature 
that gives the maximum of its  black body emission in the R-band   
is $kT\sim$ 1 eV. Equating 
$k \Teff =1$ eV in Eq. (\ref{16}) we find 
$r({\m{kT}}\sim 1$ eV)$ \sim 4.8\times 10^9 \eta^{-1/3}$ cm. Then, 
\be
\frac{\LR}{\Lx} \simeq \left(\frac{R_*}{2 r(kT= 1 {\m{eV}})}\right) 
\simeq 1\times 10^{-4}~ \eta^{-2/3}      
\label{17}      
\ee  
which is in agreement with the observed ratio reported by 
Hulleman et al. (2000) for the values of $\eta$ near unity.
To estimate the IR flux from the disk, we equate $k \Teff\simeq 0.1$ eV  
in Eq. (\ref{16}), and find 
$r(kT\simeq 0.1$ eV)$ \simeq 1\times 10^2~ r_9 \simeq 10^{11}
\eta^{-1/3}$~cm. 
Eq. (17) gives $(\LK/\Lx) \sim 10^{-5} \eta^{-2/3}$, that is, 
$\LK/\LR\sim 10^{-1}$ (independent of $\eta$)
which is also consistent  
with the observations of the AXP 4U 0142+61. 
On the other hand, shorter wavelength optical and UV emission 
which is expected from the innermost regions of the disk, could 
be effectively  absorbed due to increasing IS absorption 
with decreasing wavelength along these energies. 

Our estimates remain consistent with the observations as long as 
$\eta$ is greater than  about $\sim 30 \%$. 
Ek\c{s}i $\&$ Alpar (2003a) showed that fall-back
disks can produce AXP period and period derivatives in time scales 
consistent with their SNR ages using dipole magnetic fields with $B_*\sim
10^{12}-10^{13}$ G. They estimated  that $\eta$ could be as
low as 0.01 for some AXPs, and pointed out that $\eta$ could be higher    
if the mass loss from the disk due to propeller effect during the life
time of the AXPs and SGRs is taken into account. 
From the disk model parameters of AXP 0142+61 
(Ek\c{s}i $\&$ Alpar 2003b), we can estimate  its present 
disk mass flow rate as $\sim 3\times 10^{16}$ g s$^{-1}$, that is, 
$\eta\sim 15 \%$ which is consistent with our rough estimate within a factor 
$\sim 2 - 3$.  
More detailed numerical calculations of the optical/IR   
emission from the AXPs and SGRs will be presented in a separate work.

In present disk models of AXPs, IR and optical luminosities are assumed 
to be dominated by the X-ray irradiation of the disk by the central X-rays
(Perna, Hernquist $\&$ Narayan 2000; 
Perna $\&$ Hernquist 2000). For the observed IR emission from the 
AXP 1RXS J1708-40, these X-ray irradiated disk models give unreasonably 
large inner disk radii ($>10^{11}$ cm) and restrict the radial extend of the 
disk *$\Delta R \sim 5\times 10^{11} - 10^{12}$ cm) 
(Israel et al. 2003). Optical/IR  emission from the low-mass 
X-ray binary disks has long been known to be X-ray irradiation 
dominated, and this could be the motivation to include the X-ray 
irradiation in the AXP disk models. However, X-ray 
could be inefficient for the AXP disks due to the following 
reason. The observationally
 supported irradiation  of the outer disk of low mass X-ray binaries 
 (LMXBs) has to be indirect, very probably due to a central extended 
 hot corona (see e.g. Dubus et al. 1999), because of the self screening 
 of the disk. A central, extended hot corona can be fed by the 
 thermally unstable matter from the inner disk where the thermal 
 instabilities are most effective (Shaviv and Wehrse 1986). 
 However, the innermost disk regions 
 of AXPs (and SGRs) are cut by the dipole magnetic fields of their 
 neutron stars, which are orders of magnitude higher than the               
 magnetic fields of the NSs in LMXBs. Therefore, the X-ray irradiation 
 that prevails at the LMXB disks, does not necessarily be effective 
 for the AXP and SGR disks, though it is not impossible.    
 Therefore, in our calculations we ignore the X-ray irradiation. 
Observed IR/optical luminosity of AXPs do not restrict the outer disk 
radius in our explanation. That is, the outer disk expands freely in the 
low viscosity state, and needs not to be passive (no viscosity) as proposed
by Menou et al. (2001). On the other hand, we expect that the inner disk
radius is around the Alfv$\acute{e}$n radius ($\sim 5\times 10^{8}$ cm) 
which is about an order of magnitude less than the estimated radial distance 
at which most of the optical R-band luminosity is emitted 
($\sim 5\times 10^{9}$ cm). Inner rim of the disk can be  
irradiated by the X-rays, but the resultant emission is expected to be in the
UV band which is easily absorbed by the ISM. It is hard to estimate the
detailed emission spectrum from the innermost disk due to the uncertainties 
in the physics of the disk-magnetosphere interaction.   
Part of the inner disk matter outflowing due to the propeller effect may 
also modify the emission from the underlying innermost disk where higher
energy optical photons emitted.  

\section{Conclusions and Discussion }

We have shown that : ($i$) the observed optical {\it pulsed} 
emission from the AXP 4U 0142+61 can be explained by means of  
both the magnetar outer gap model and the disk-star 
dynamo gap model, and therefore does not eliminate either of the
models,  
$(ii)$ the two models, which do not provide an observational test
through their different synchrotron X-ray luminosity due to the
dominating X-ray luminosity from the NS, can be tested observationally 
via their estimated high energy gamma-ray spectra  
 by the Gamma-ray Large-Area Space Telescope (GLAST),  
$(iii)$ the IR and optical emission  expected 
from a standard accretion disk  model 
is in  agreement with the reported unpulsed IR and optical 
luminosity of the AXP 4U 0142+61 for the ratio of the 
accretion rate onto the NS to the disk mass flow rate 
$\eta~ \ga~ 30 \%$. 

Our estimation for $\eta$ is roughly consistent with that  
obtained   from the fall-back disk model of AXP 0142+61 
(Ek\c{s}i $\&$ Alpar 2003b).
The IR to X-ray luminosity ratio
of the AXP 1E 2259+58  is similar to that of  4U 0142+61
(Hulleman 2001), and can be explained by similar arguments.
The reported $\LR/\Lx$ ratio of AXP 1RXS J1708-40
is $\sim 2\times 10^{-3}$ (Israel et al. 2003) which is about 10 times
higher than that of 4U 0142+61, and might indicate
relatively more efficient mass outflow due to the propeller effect in
this system with $\eta \sim 1/30$. Detiled numerical 
disk calculations to estimate the optical/IR luminosities of AXPs and SGRs 
and comparisons with their period and period evolution models  
will be presented in future work.

\acknowledgments
We thank Ali Alpar and Yavuz Ek\c{s}i for discussions, 
and comments on the manuscript. \"{U}.E. acnowledges support 
from the High Energy Astrophysics Research Group TBAG-\c{C}-4 
of T\"{U}B{\.I}TAK and from the Astrophysics and Space Forum at 
Sabanc{\i} University. 
Part of this work was supported by RGC grant of Hong Kong 
Government.


\clearpage
\begin{figure}
\plotone{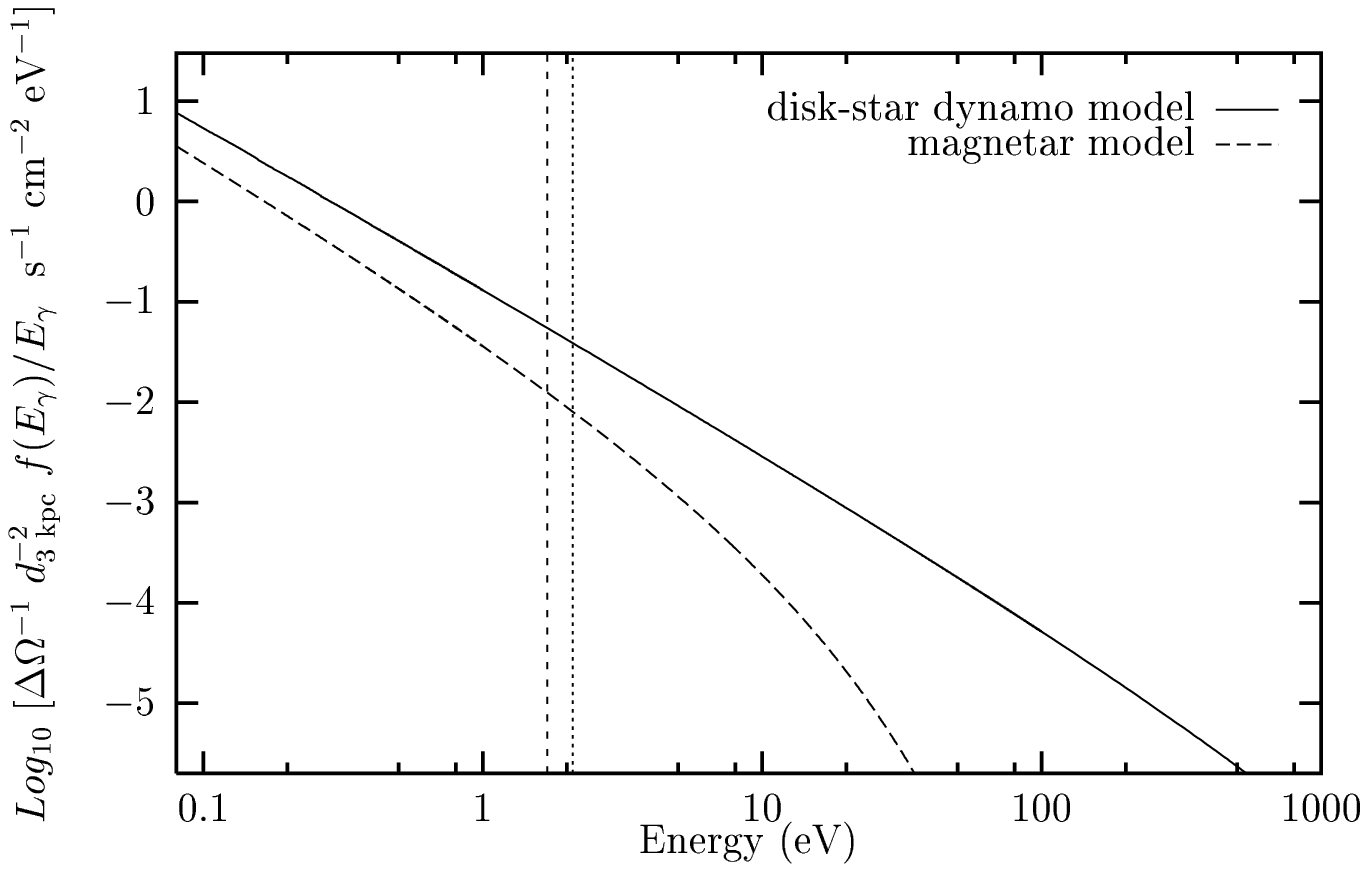}
\vspace{-7cm}
\caption{Photon flux  spectrum of the  synchrotron radiation  from
the magnetar outer gap
model (dashed line) and the disk-star dynamo model
(continuous line). 
The vertical lines show the borders of the R-band 
(1.7 - 2.1~ eV).} 

\end{figure}

\end{document}